\documentclass[aps,prl,twocolumn,fleqn,floatfix,superscriptaddress]{revtex4}

\usepackage{epsfig,color}
\usepackage{amsmath,amssymb,amsfonts}

\newcommand \eg {{e.~g.}}

\begin{document}

\title{Small Shear Viscosity of a Quark-Gluon Plasma Implies Strong Jet Quenching}

\author{Abhijit Majumder}
\affiliation{Department of Physics, Duke University, Durham, NC 27708}
\author{Berndt M\"uller}
\affiliation{Department of Physics, Duke University, Durham, NC 27708}
\author{Xin-Nian Wang}
\affiliation{Nuclear Science Division, MS 70R0319, Lawrence Berkeley National 
                   Laboratory, Berkeley, CA 94720}

\date{\today}

\begin{abstract}
We derive an expression relating the transport parameter $\hat{q}$ 
and the shear viscosity $\eta$ of a weakly coupled quark-gluon plasma. 
A deviation from this relation can be regarded as a quantitative measure 
of ``strong coupling'' of the medium. The ratio $T^3/\hat{q}$, where 
$T$ is the temperature, is a more broadly valid measure of the coupling 
strength of the medium than $\eta/s$, where $s$ denotes the entropy 
density. Different estimates of $\hat{q}$ derived from existing RHIC data 
are shown to imply radically different structures of the produced matter. 
\end{abstract}

\maketitle

The highly excited, strongly interacting matter formed in collisions of large
nuclei at the Relativistic Heavy Ion Collider (RHIC) exhibits two unusual 
properties:  The emission of hadrons with large transverse momentum is strongly 
suppressed in central collisions~\cite{jet-quenching} and 
the collective flow of the matter is well described by relativistic hydrodynamics 
with a negligible shear viscosity~\cite{Teaney:2003kp}.
The suppression of hadrons at large transverse momentum, generally referred 
to as jet quenching, is understood to be caused by gluon radiation induced by 
multiple collisions of the leading parton with color charges in the near-thermal 
medium \cite{Baier:1996kr,energy-loss,GuoW}. 
This process is governed by a transport 
parameter $\hat{q}$, which is the squared average transverse momentum exchange 
between the medium and the fast parton per unit path length~\cite{Baier:1996kr}. 

The RHIC data have been interpreted to imply that the quark-gluon plasma 
produced in the nuclear collision is a strongly coupled medium, which may not
even contain quasi-particles whose interactions can be treated in some effective 
perturbation theory~\cite{Gyulassy:2004zy}. 
While the absolute values of the shear viscosity-to-entropy ratio
$\eta/s$ determined from model analysis of 
RHIC data tend to rule out a class of weak coupling approaches based on the 
Hard-Thermal Loop (HTL) approximation, it may still be consistent with the general 
picture based on weakly coupled quasi-particles with partonic (quark and gluon) 
quantum numbers. The present letter outlines a method for quantitatively testing 
this hypothesis.

The interaction of hard jets with strongly interacting matter may always be treated 
in a perturbative expansion involving short lived partonic excitations as the QCD 
coupling is weak at short distances. Thermal excitations in a QCD medium, on the 
other hand,  may or may not be partonic quasi-particles. Here, we take the term 
``weakly coupled'' to mean that the properties of the medium can be described 
perturbatively on the basis of an appropriate, partonic quasi-particle picture. 
This notion does not preclude the possibility that the quasi-particles themselves may 
emerge nonperturbatively from the fundamental quanta of QCD. ``Strong coupling'' 
indicates the absence of such a partonic quasi-particle description. The central 
assertion advanced here is that there exists a general relation between the jet parameter 
$\hat{q}$ and the coefficient of shear viscosity $\eta$ which holds for any weakly 
coupled partonic plasma where the interaction between the quasiparticles, which is 
responsible for the generation of a viscosity, has the same structure and strength as 
the interaction between the leading jet parton and the medium.  

In any partonic quasi-particle framework, the transport parameter governing the 
radiative energy loss of a propagating parton in SU(3)-color representation 
$R$ is~\cite{Baier:1996kr}:
\begin{equation}
\hat{q}_R \equiv 
   \rho \int dq_\perp^2\, q_\perp^2 \frac{d\sigma_R}{dq_\perp^2} .
\label{eq:qhat-sigma}
\end{equation}
The shear viscosity of a fluid is defined as the coefficient of the contribution 
to its stress tensor, which is proportional to the divergence-free part of the 
velocity gradient. In the framework of kinetic theory based on a quasi-particle
picture, the shear viscosity $\eta$ is determined by the mean-free path 
$\lambda_{\rm f}(p)$ of a constituent particle of momentum $p$ in the medium:
\begin{equation}
\eta  \sim C \rho \langle p \rangle \lambda_{\rm f} ,
\label{eq:visc}
\end{equation}
with $C \approx 1/3$ \cite{Danielewicz:1984ww,deGroot}.
A heuristic connection between $\eta$ and $\hat{q}$ can be established by 
the observation that the mean-free path is related to the average transport 
cross section of a quasi-particle in the medium: 
$\lambda_{\rm f} = (\rho\,\sigma_{\rm tr})^{-1}$. When soft scattering 
dominates, as in the case of perturbative QCD, the transport cross section 
is related to the differential cross section by the relation:
\begin{equation}
\sigma_{\rm tr} \approx \frac{4}{\hat{s}}  
   \int dq_{\perp}^2\, q_{\perp}^2\, \frac{d\sigma}{dq_{\perp}^2} 
\equiv \frac{4\hat{q}}{\hat{s}\rho}  ,
\label{eq:sigma-tr}
\end{equation}
where $\sqrt{\hat{s}}$ is the center-of-mass energy. For a thermal 
ensemble of massless particles, $\langle p \rangle \approx 3T$ and 
$\langle \hat{s} \rangle \approx 18 T^2$, and thus:
\begin{equation}
\eta \approx 13.5\, C \frac{T^3\rho}{\hat{q}} .
\label{eq:eta-qhat-approx}
\end{equation}
Using the relation $s \approx 3.6\,\rho$ for the entropy density of a gas of free,
massless bosons, the following heuristic equation relating the ratio of the shear 
viscosity to the entropy density with the transport coefficient may be derived:
\begin{equation}
\frac{\eta}{s} \approx 3.75\, C \frac{T^3}{\hat{q}} .
\label{eq:eta-s1}
\end{equation}
This relation shows that a large value of $\hat{q}$ implies a small value 
for the ratio $\eta/s$, which is thought to be bounded by the quantum 
limit $\eta/s \geq (4\pi)^{-1}$~\cite{Kovtun:2004de}. 

When perturbation theory can be applied to the medium, the jet quenching
parameter can  be expressed in terms of the light-cone gluon correlation 
function
\begin{equation}
\hat{q}_R =\frac{4\pi C_R \alpha_{\rm s}}{N_c^2-1} \int dy^-
\left\langle F^{ai+}(0)F_{i}^{a+}(y^-) \right\rangle e^{i \xi p^+y^-}.
\label{eq:qhat}
\end{equation}
In Eq.~(\ref{eq:qhat}), 
$\langle {\cal O} \rangle =(2\pi)^{-3} \int d^3p/2p^+ f(p) \langle p| {\cal O} |p\rangle$
denotes the ensemble average of an operator ${\cal O}$ in the medium
composed of states $|p\rangle$ with occupation probability $f(p)$,
$\xi=\langle k_T^2\rangle/2E\langle p^+\rangle$, $\langle k_T^2\rangle$ 
is the average transverse momentum carried by the gluons in $|p\rangle$, 
and the index $i$ sums over the two transverse  directions. 
$\rho=\int d^3p f(p)/(2\pi)^3$ denotes the density of scattering centers, 
mainly gluons, in the matter. $d\sigma_R/dq_{\perp}^2$ is the differential 
cross section for a parton on a scattering center. A nonperturbative definition 
of $\hat{q}_R$ in terms of a Wilson loop along the light-cone has been proposed
in ref.~\cite{Liu:2006ug}. 

In principle, $\hat{q}_R(E)$ depends weakly on the energy $E$ and virtuality of the 
jet-inducing parton. In the limit of a thick (large opacity) medium of thickness $L$, the 
virtuality of the parton is determined by the saturation scale $Q_s$ of the medium
\cite{Liu:2006ug}, and the transport parameter for parton energy loss takes on the 
universal form $\hat{q}=Q_s^2/L$. We will neglect a possible logarithmic energy 
dependence of $\hat{q}$ here. For a massless parton in a weakly coupled gluon 
plasma:
\begin{equation}
\hat{q}_R = \frac{8\zeta(3)}{\pi} N_c C_R \alpha_{\rm s}^2 T^3 \ln\frac{1}{\alpha_s},
\label{eq:qhatBDMPS}
\end{equation}
in the leading log approximation~\cite{Baier:2006fr}. In the remainder of this paper, 
we will omit the subscript in the notation for the transport parameter 
$\hat{q}\equiv \hat{q}_A$ for a gluon jet.

The expression for the shear viscosity depends on the mechanism that limits
momentum transport in the medium. We start with the shear viscosity due to parton 
collisions in the quark-gluon plasma, following Arnold et al. \cite{Arnold:2000dr}. 
In the Chapman-Enskog approach to linearized transport theory, one 
parametrizes the deviation $f_1(p)$ of the parton distribution from equilibrium 
due by a function $\bar\Delta(p)$ in the form (using the notation 
of~\cite{Asakawa:2006})
\begin{equation}
f_1(p) = - \frac{\bar\Delta(p)}{ET^2} p_ip_j (\nabla u)_{ij} ,
\label{eq:f1}
\end{equation}
where $(\nabla u)_{ij}$ denotes the traceless velocity gradient. 
One can then derive an integral equation for $\bar\Delta(p)$ from the 
linearized Boltzmann equation. An analytic estimate can be obtained by 
restricting $\bar\Delta$ to the functional form $\bar\Delta(p) = Ap/T$. 
The shear viscosity is expressed in terms of $\bar\Delta(p)$ as
\begin{equation}
\eta_{\rm C} = - \frac{1}{15\, T} \int \frac{d^3p}{(2\pi)^3} \frac{p^4}{E^2}
   \bar\Delta(p) \frac{\partial f_0}{\partial E} \, ,
\label{eq:eta-Delta}
\end{equation}
where $f_0(p)$ denotes the equilibrium distribution of partons. Assuming 
that collisions in the medium are dominated by soft scattering, the 
integral over the differential scattering cross section can be extracted from
the collision integral, yielding a factor $\hat{q}/\rho$, where $\hat{q}$ is
defined by Eq.~\eqref{eq:qhat-sigma} for a gluon jet. After a lengthy 
calculation one finds that in this limit the shear viscosity for a thermal gluon
 plasma  can be expressed in the form:
\begin{equation}
\eta_{\rm C} \approx C' \, \frac{T^3\rho}{\hat{q}} .
\label{eq:eta-c2}
\end{equation}
In order to extract the constant, we can make use of the known result for the 
shear viscosity of a pure gluon gas in the leading logarithmic approximation 
\cite{Arnold:2000dr}: 
\begin{equation}
\eta_{\rm LL} = \frac{3.81}{\pi^2}  \frac{N_c^2-1}{N_c^2}
\frac{T^3}{\alpha_s^2\ln(1/\alpha_s)} . 
\label{eq:eta-amy}
\end{equation}
Inserting the jet quenching parameter in the weak coupling limit 
\eqref{eq:qhatBDMPS} and using the expression for the density of a 
thermal gas of free gluons, we obtain $C' \approx 4.85$. Comparing with 
Eq.~\eqref{eq:eta-qhat-approx}, we find that this corresponds to
$C \approx 0.36$.

We next consider the case of anomalous shear viscosity,
which is generated by dynamically created turbulent color fields in a 
rapidly expanding quark-gluon plasma \cite{Asakawa:2006}. 
Unstable collective plasma modes \cite{instabilities} result in the growth of long 
wave length modes of the glue fields, which ultimately saturate due to their 
nonlinear self-interaction \cite{Rebhan:2005re,Arnold:2005ef}. The turbulent 
quark-gluon plasma is characterized by a non-vanishing expectation value of 
the gluon correlator $\langle F^{\sigma +}(0)F_{\sigma}^{+}(y^-) \rangle$ of the 
soft color fields.

The presence of localized domains of color fields induces an anomalous 
contribution to the shear viscosity of the matter
\cite{Asakawa:2006}. This contribution is derived by 
considering the effect of random color fields on the propagation of quasi-thermal 
partons, which is described by a Fokker-Planck equation of the form:
\begin{equation}
\left[ \frac{\partial}{\partial t} + {\mathbf v}\cdot\nabla_r 
- \nabla_p D^R({\mathbf p},t) \nabla_p \right] f({\mathbf p},{\mathbf r},t) = 0
\label{eq:FP}
\end{equation}
with the average parton phase space distribution $f$ and the diffusion tensor
\begin{equation}
D_{ij}^R = \int_{-\infty}^t dt' 
\left\langle F_i(\bar{r}(t'),t') F_j(\bar{r}(t),t) \right\rangle .
\label{eq:D-def}
\end{equation}
Here ${\mathbf F}=g Q_a^R ({\mathbf E}^a + {\mathbf v}\times{\mathbf B}^a)$ 
is the color Lorentz force generated by the turbulent color fields and $Q^R$
is color charge of the parton in the given representation. The anomalous shear 
viscosity implied by Eq.~\eqref{eq:FP} can be evaluated, \eg, for randomly 
distributed color fields with a correlation time $\tau_{\rm m}$ along the light-cone
\cite{Asakawa:2006}:
\begin{equation}
\eta_{\rm A} = \frac{16\,\zeta(6)(N_c^2-1)^2T^6}
                                 {\pi^2N_c g^2 \langle E^2+B^2 \rangle \tau_{\rm m}} .
\label{eq:eta-A}
\end{equation}

The diffusion coefficient $D_{ij}$ in Eq.~\eqref{eq:D-def}, is closely related 
to the transport parameter for radiative energy loss, Eq.~\eqref{eq:qhat}.
Applying the definition of $\hat{q}_A$ as rate of growth of the transverse 
momentum fluctuations of a fast gluon to an ensemble of turbulent color fields, 
one finds (for partons in the adjoint color
representation):
\begin{equation}
\hat{q} = \frac{8\pi\alpha_sN_c}{3(N_c^2-1)}\,  \langle E^2+B^2 \rangle \tau_{\rm m} .
\label{eq:qhat-EB}
\end{equation}
Combining this expression with Eq.~\eqref{eq:eta-A} we obtain a relation 
between the contributions of turbulent color fields to the transport parameter 
in gluon energy loss and the anomalous shear viscosity of gluons:
\begin{equation}
\eta_{\rm A} = \frac{32\zeta(6)(N_c^2-1)}{3\pi^2} \frac{T^6}{\hat{q}} 
\approx 4.51\, \frac{T^3\rho}{\hat{q}} .
\label{eq:eta-qhat}
\end{equation}
The numerical coefficient here is very close to the one obtained for
the collisional viscosity. Alternatively, we can calculate the dimensionless 
ratio of the shear viscosity and the entropy density. 
For a free thermal gluon gas, the entropy density is given by
\begin{equation}
s = 2(N_c^2-1)\, \frac{2\pi^2}{45} T^3 ,
\label{eq:s}
\end{equation}
which implies the following result for the shear viscosity-to-entropy ratio:
\begin{equation}
\frac{\eta_{\rm A}}{s} = \frac{8\pi^2}{63} \frac{T^3}{\hat{q}} 
\approx 1.25\,  \frac{T^3}{\hat{q}} ,
\label{eq:eta/s}
\end{equation}
again in agreement with \eqref{eq:eta-s1} for $C \approx 1/3$. This confirms our
assertion that Eq.~\eqref{eq:eta-s1} holds if the transport properties of a 
quark-gluon plasma are described by kinetic theory based on partonic 
quasi-particles.

In order to explore the opposite limit, we first consider a solvable strongly coupled theory, 
$N=4$ supersymmetric Yang-Mills (SYM) theory. Analytical results for $\eta/s$ and 
$\hat{q}$ have been derived for this theory in the limit of large 't Hooft coupling 
$\lambda = g^2N_c$. The strong coupling expression for the ratio of shear viscosity
to entropy density is well established \cite{Policastro:2001yc,Buchel:2004di}:
\begin{equation}
\frac{\eta}{s} = \frac{1}{4\pi} 
   \left[ 1+\frac{135\,\zeta(3)}{(8\lambda)^{3/2}} + \cdots \right] .
\label{eq:eta-sym}
\end{equation}
There is not yet universal agreement about the correct generalization of the jet 
quenching parameter. Here we follow Liu {\em et al.} \cite{Liu:2006ug} who 
defined $\hat{q}$ by means of an adjoint Wilson loop along the light-cone 
and found the result
\begin{equation}
\hat{q} = \frac{\pi^{3/2}\Gamma(3/4)}{\Gamma(5/4)} \sqrt{\lambda} T^3
  \approx 7.53\,\sqrt{\lambda} T^3 .
\label{eq:qhat-sym}
\end{equation}
Inserting this expression into the right-hand side of \eqref{eq:eta-s1} for 
$C \approx 1/3$ we obtain
\begin{equation}
1.25\, \frac{T^3}{\hat{q}} \approx \frac{0.166}{\sqrt{\lambda}}
\ll \frac{\eta}{s}  ,
\label{eq:rhs}
\end{equation}
for $\lambda \gg 4.35$. Obviously, the left-hand side of \eqref{eq:rhs} is a better 
measure of the strength of the coupling in the strong coupling limit than the 
right-hand side.

The strong coupling limit of QCD exhibits quark confinement. As is well known,
this limit of QCD is also amenable to rigorous calculations in the chiral limit and
at low temperature. In this limit, the shear viscosity-to-entropy density ratio of
a pion gas is given by \cite{Chen:2006ig}:
\begin{equation}
\frac{\eta}{s} \approx \frac{15f_\pi^4}{16\pi T^4}
\label{eq:etas-pi}
\end{equation}
The ratio grows rapidly with decreasing temperature, because the pion-pion
interaction weakens at low momentum. On the other hand, the jet quenching 
parameter for a gluon jet propagating through a pion gas has the form
\begin{equation}
\hat{q}_A^{(\pi)} \approx 
\frac{4\pi^2N_c\alpha_s}{N_c^2-1} \rho_\pi [x G_\pi(x)]_{x\to 0} ,
\label{eq:qhat-pi}
\end{equation}
where $G_\pi(x)$ is the gluon distribution function of the pion.
Since the pion density in the chiral limit is proportional to $T^3$, this implies
that the left-hand side of \eqref{eq:rhs} is approximately independent of $T$:
\begin{equation}
1.25 \frac{T^3}{\hat{q}_A^{(\pi)}} \approx \frac{0.235}{\alpha_s [x G_\pi(x)]_{x\to 0}} .
\label{eq:T3qhat-pi}
\end{equation}
Again, $T^3/\hat{q}$ is a much better measure of the underlying QCD coupling 
strength in this domain than $\eta/s$. We note that $T^3/\hat{q} \ll \eta/s$
in both cases of strong coupling. It is not surprising that the heuristic relation 
\eqref{eq:eta-qhat-approx}  does not hold in the strong coupling limit: 
In a strongly coupled system, thermal modes are not approximately described 
as elementary quasi-particles \cite{Teaney} probed by jet quenching. 
On the other hand, highly energetic 
excitations can retain a quasi-particle nature, and their interactions with the 
medium continue to track the strength of the coupling. The violation of the relation 
\eqref{eq:eta-qhat-approx} may thus be considered as a general criterion for 
``strongly coupling'' in  a gauge theory.

We can now ask whether the relation \eqref{eq:eta-qhat-approx} 
applies to the matter produced at RHIC, according to the phenomenological 
analysis of jet quenching measurements. At present, the values of $\hat{q}$ 
deduced from the data by means of different approaches to jet quenching  
differ from each other by substantial factors~\cite{Majumder:2007iu}. Recent 
studies within the framework of the twist expansion, fitting 
to the experimental data on hadron suppression in the most central Au+Au
collisions at RHIC~\cite{Zhang:2007ja,Majumder:2007ae} gave values
$\hat{q}_0^{\rm (ht)} = 1 - 2 \; {\rm GeV}^2/{\rm fm}$ 
for the gluon quenching parameter at the initial time $\tau_0 \approx 1$ fm/c. 
A study using a variant of the eikonal approach~\cite{Baier:1996kr} 
resulted in the estimate~\cite{Sahlmueller:2007wx}
$\hat{q}_0^{\rm (eik)} = 10 - 30 \; {\rm GeV}^2/{\rm fm}$ 
for the same quantity (scaled from the value given in~\cite{Sahlmueller:2007wx} 
for a propagating quark). As we now show, the two different values imply 
radically different properties of the medium.

One can estimate \cite{Muller:2005en} the average initial entropy density 
from the measured charge hadron multiplicity of the most central Au+Au 
collisions as $s_0=(33\pm 3)$ fm$^{-3}$ at $\tau_0=1$ fm/c corresponding 
to an initial temperature $T_0 = (337\pm 10)$ MeV. Using Eq.~(\ref{eq:eta-s1}) 
with $C=1/3$ one then finds that the lower value $\hat{q}_0^{\rm (ht)}$ implies
$1.25\, T_0^3/\hat{q}_{0} = 0.12 - 0.24$, whereas the higher value
$\hat{q}_0^{\rm (eik)}$ implies $1.25\, T_0^3/\hat{q}_{0} = 0.008 - 0.024$.
This first result is close to the conjectured lower bound for $\eta/s$ and lies well 
within the range of values for the shear viscosity-to-entropy density ratio 
($\eta/s < 0.3$) which are compatible with the measured hadron 
spectra from Au+Au collisions at RHIC \cite{Baier:2006gy}. If ultimately 
confirmed, it would indicate that the quark-gluon plasma produced in Au+Au
collisions is (marginally) weakly coupled. On the other hand, the second 
result lies far below the lower bound for $\eta/s$ and thus implies that the
quark-gluon plasma formed at RHIC is deep in the strong coupling regime
and cannot be described as a quasi-particle plasma.

In summary,  we have derived a general relation between the shear viscosity 
$\eta$ and the jet quenching parameter $\hat{q}$ for a ``weakly coupled'',
{\em i.e.} quasi-particle dominated quark-gluon plasma. The relation associates a 
small ratio of shear viscosity to entropy density to a large value of the jet 
quenching parameter. 

The fact that $\eta/s$ saturates in the limit of strong coupling of the SYM theory 
but $\hat{q}$ continues to increase, suggests that the ratio $T^3/\hat{q}$ may 
serve as a more broadly applicable measure of the coupling strength of a 
quark-gluon plasma. We thus conjecture that the following relations hold 
generally:
\begin{equation}
\frac{\eta}{s}
\left\{ \begin{array}{c}
\approx \\
\gg
\end{array} \right\}
1.25\, \frac{T^3}{\hat{q}} \qquad 
\left\{ \begin{array}{l}
{\rm for~weak~coupling,} \\
{\rm for~strong~coupling.} 
\end{array} \right.
\label{eq:criterion}
\end{equation}
An unambiguous determination of both sides of \eqref{eq:criterion} 
from experimental data would thus permit a model independent, quantitative 
assessment of the strongly coupled nature of the quark-gluon plasma produced 
in heavy ion collisions \cite{Blaizot:2007}.

{\it Acknowledgments:} This work was supported in part by the 
U.~S.~Department of Energy under grant DE-FG02-05ER41367 and
contract DE-AC02-05CH11231. BM and XNW acknowledge the hospitality
of the Galileo Galilei Institute (Firenze) where this manuscript was completed.
We thank K.~Rajagopal for valuable comments.


\begin{thebibliography}{99}

\bibitem{jet-quenching}
  K.~Adcox {\it et al.}, 
  Phys.\ Rev.\ Lett.\  {\bf 88}, 022301 (2002);
  C.~Adler {\it et al.}, 
  Phys.\ Rev.\ Lett.\ {\bf 89} 202301 (2002).

\bibitem{Teaney:2003kp}
  D.~Teaney,
  Phys.\ Rev.\ C {\bf 68}, 034913 (2003).
  
\bibitem{energy-loss}
  M.~Gyulassy and X.-N.~Wang,
  Nucl.\ Phys.\ {\bf 420} 583 (1994);
  B. G. Zhakharov, 
  JETP Lett.\ {\bf 63}, 952 (1996);
  U. Wiedemann, 
  Nucl. Phys. {\bf B588}, 303 (2000).
  M. Gyulassy, P. L\'evai and I. Vitev, 
  Nucl.\ Phys.\ {\bf B594}, 371 (2001).

\bibitem{Baier:1996kr}
  R.~Baier, Y.~L.~Dokshitzer, A.~H.~Mueller, S.~Peigne and D.~Schiff,
  Nucl.\ Phys.\ B {\bf 483} 291 (1997).

\bibitem{Gyulassy:2004zy}
  M.~Gyulassy and L.~McLerran 2005.
  {\em Nucl.\ Phys.\ A} 750:30. 
  
\bibitem{GuoW} 
  X.~F.~Guo and X.-N.~Wang,
  Phys.\ Rev.\ Lett.\  {\bf 85}, 3591 (2000);
  Nucl.\ Phys.\ A {\bf 696}, 788 (2001).

\bibitem{Danielewicz:1984ww}
  P.~Danielewicz and M.~Gyulassy,
  Phys.\ Rev.\  D {\bf 31}, 53 (1985).
  
\bibitem{deGroot}
  S.~R.~deGroot, W.~A.~van Leeuwen, and Ch.~G.~van Weert,
  {\em Relativistic Kinetic Theory},
  (North-Holland, Amsterdam, 1980).

\bibitem{Kovtun:2004de}
  P.~Kovtun, D.~T.~Son and A.~O.~Starinets,
  Phys.\ Rev.\ Lett.\  {\bf 94}, 111601 (2005).

\bibitem{Liu:2006ug}
  H.~Liu, K.~Rajagopal and U.~A.~Wiedemann,
  Phys.\ Rev.\ Lett.\  {\bf 97}, 182301 (2006);
  arXiv:hep-ph/0612168.


\bibitem{Baier:2006fr}
  R.~Baier and D.~Schiff,
  JHEP {\bf 0609}, 059 (2006).

\bibitem{Arnold:2000dr}
  P.~Arnold, G.~D.~Moore and L.~G.~Yaffe,
  JHEP {\bf 0011}, 001 (2000)
  [arXiv:hep-ph/0010177].

\bibitem{Asakawa:2006}
 M.~Asakawa, S.~A.~Bass and B.~M\"uller,
  Phys.\ Rev.\ Lett.\  {\bf 96}, 252301 (2006);
  Prog.\ Theor.\ Phys. {\bf 116}, 725 (2006).
 
\bibitem{instabilities}    
  J.~Randrup and S.~Mrowczynski,
  Phys.\ Rev.\ C {\bf 68}, 034909 (2003);
  P.~Romatschke and M.~Strickland,
  Phys.\ Rev.\ D {\bf 68}, 036004 (2003);
  P.~Arnold, J.~Lenaghan and G.~D.~Moore,
  JHEP {\bf 0308}, 002 (2003).
  
\bibitem{Rebhan:2005re}
  A.~Rebhan, P.~Romatschke and M.~Strickland,
  JHEP {\bf 0509}, 041 (2005).
  
\bibitem{Arnold:2005ef}
  P.~Arnold and G.~D.~Moore,
  Phys.\ Rev.\ D {\bf 73}, 025006 (2006);
  Phys.\ Rev.\ D {\bf 73}, 025013 (2006).

\bibitem{Majumder:2006wi}
  A.~Majumder, B.~M\"uller and S.~A.~Bass,
  arXiv:hep-ph/0611135.

\bibitem{Policastro:2001yc}
  G.~Policastro, D.~T.~Son and A.~O.~Starinets,
  Phys.\ Rev.\ Lett.\  {\bf 87}, 081601 (2001).
  
\bibitem{Buchel:2004di}
  A.~Buchel, J.~T.~Liu and A.~O.~Starinets,
  Nucl.\ Phys.\  B {\bf 707}, 56 (2005).

\bibitem{Teaney}
  D.~Teaney,
  Phys.\ Rev.\  D {\bf 74}, 045025 (2006).
    
  
\bibitem{Herzog:2006gh}
  C.~P.~Herzog, A.~Karch, P.~Kovtun, C.~Kozcaz and L.~G.~Yaffe,
  JHEP {\bf 0607}, 013 (2006).

\bibitem{Chen:2006ig}
  J.~W.~Chen and E.~Nakano,
  Phys.\ Lett.\  B {\bf 647}, 371 (2007).
  
\bibitem{Wang:2000uj}
  X.~N.~Wang,
  Phys.\ Lett.\ B {\bf 485}, 157 (2000).

\bibitem{Dokshitzer:2001zm}
  Y.~L.~Dokshitzer and D.~E.~Kharzeev,
  Phys.\ Lett.\  B {\bf 519}, 199 (2001).

\bibitem{Zhang:2007ja}
  H.~Zhang, J.~F.~Owens, E.~Wang and X.~N.~Wang,
  arXiv:nucl-th/0701045.

\bibitem{Majumder:2007ae}
  A.~Majumder, C.~Nonaka and S.~A.~Bass,
  arXiv:nucl-th/0703019.
  
\bibitem{Sahlmueller:2007wx}
  B.~Sahlmueller (PHENIX collaboration),
  arXiv:nucl-ex/0701060.
  
\bibitem{Muller:2005en}
  B.~M\"uller and K.~Rajagopal,
  Eur.\ Phys.\ J.\ C {\bf 43}, 15 (2005).

\bibitem{Baier:2006gy}
  R.~Baier and P.~Romatschke,
  arXiv:nucl-th/0610108.
  
\bibitem{Majumder:2007iu}
  A.~Majumder,
  arXiv:nucl-th/0702066.

\bibitem{Blaizot:2007}
  J.-P.~Blaizot,
  arxiv:hep-ph/0703150.

\end{thebibliography}
\end{document}